# Atom probe microscopy investigation of Mg site occupancy within δ′ precipitates in an Al-Mg-Li alloy

Baptiste Gault[1,2], Xiang Yuan Cui[1], Michael P. Moody[1], Frederic de Geuser[3], Christophe Sigli[4], Simon P. Ringer[1,5], Alexis Deschamps[3]

[1]*Australian Centre for Microscopy and Microanalysis, Madsen Building F09, The University of Sydney, NSW 2006, Australia*

[2]*Institute of Materials and Engineering Science, Australian Nuclear Science and Technology Organisation, Private Mail Bag 1, Menai, NSW 2234, Australia*

[3]*SIMAP, INP Grenoble – CNRS – UJF, BP 75, 38402 St Martin d'Hères Cedex, France*

[4] *Constellium – Centre de Recherches de Voreppe, BP 27, 38341 Voreppe Cedex, France*

[5] *ARC Centre of Excellence for Design in Light Metals, The University of Sydney, NSW 2006, Australia*

## ABSTRACT

The composition and site-occupancy of Mg within ordered $\delta'$ precipitates in a model Al–Mg–Li alloy have been characterized by atom probe microscopy and first-principles simulations. The concentration in the precipitates is found to be almost the same as that of the matrix, however, we show evidence that Mg partitions to the sites normally occupied by Li in the $L1_2$ structure. Density functional calculations demonstrate that this partition is energetically favorable, in agreement with experimental results.

* Corresponding author: baptiste.gault@sydney.edu.au - Tel: + 61 2 93517548

In recent years there has been a renaissance in the research and development of Al-Li-based alloys, driven by their great potential for weight reduction in structural aerospace applications. In this context, Al–Li–Mg alloys (1420 alloy series) are of fundamental interest because they are lightweight, weldable, corrosion resistant and exhibit moderate strength [1, 2]. It has previously been shown that, after solution treatment, a fast quench to room temperature and thermal ageing at intermediate temperatures (100-170°C), these Al–Li–Mg alloys exhibit a microstructure dominated by the metastable δ′ phase [1, 3]. This phase, sometimes referred to as a GP-zone [4], has an expected stoichiometry of $Al_3(Li, Mg)$, is fully coherent with the FCC-Al matrix, possesses an ordered $L1_2$ structure, and appears very similar to the spherical $Al_3Li$ precipitates found in the Al-Li binary system [5, 6].

Over the years, the precipitation of the δ′ phase in Al–Mg–Li alloys has been the focus of many investigations, and yet much remains unknown about this phase transformation [3, 4, 7, 8]. Of particular interest is the exact role of Mg, which seems to greatly enhance the precipitation of δ′ in comparison to that observed in the Al–Li binary system. Indeed, Mg has a high solubility in Al and precipitation of $Al_3Mg$ is not expected in the compositional range commonly used in these alloys (1–5.5 at. %). A more detailed understanding of the atomic scale mechanisms that control the precipitation process requires the precise measurement of the Mg concentration within the precipitates. More specifically, the challenge is to explore whether there occurs a partitioning of Mg atoms within the δ′ $L1_2$ lattice structure on to either the Li or Al sub-lattices, or otherwise. The experimental observation that high Al alloys that contain high levels of Mg (>10 at. %) can precipitate a metastable $L1_2$ $Al_3Mg$ phase (β′′) would suggest that a preferential partitioning of Mg occurs on the Li sub-lattice [9, 10]. However, the large volume fraction of precipitates, the low atomic number of Li and the small difference in atomic number between Al and Mg, conspire to make observations of this partitioning extremely challenging and not obviously amenable to conventional scattering techniques.

Here, we have applied atom probe tomography (APT) to characterize the distribution of Mg within a laboratory alloy of composition Al–5.18Mg –6.79Li (at.%) (Al–5Mg–1.8Li (wt. %)) aged for 8h at 150°C and also after 24h at 120°C. APT has the unique capacity for simultaneous chemical identification and precise three-dimensional location of the individual atoms in a material [11]. In particular, it has been demonstrated that extremely high spatial resolution can be achieved in the analysis of Al and its alloys [12]. APT is based upon the effect of an intense electric field to induce the ionization and desorption of individual atoms from the surface of a needle-shaped specimen subjected to the superposition of a DC high-voltage (HV) and fast HV-pulses [13]. This process is known as field evaporation. The field emitted ions are collected by a position-sensitive detector. Their time-of-flight is recorded and translated into a mass-to-charge-state ratio and the positional information is sequentially analysed to build a tomographic reconstruction of the field-evaporated volume [14].

It is well documented that crystallographic information is partially retained within atom probe reconstructions and this is exemplified by the capacity of the technique to directly image atomic planes [15]. Several methods have previously been developed to interrogate the atomic-scale crystallographic information within these tomograms, including: Fourier transform (FT) approaches [16, 17], three-dimensional autocorrelation functions, termed spatial distribution maps (SDM), which are usually plotted in the form of a 1D-histogram along a specific crystallographic direction [18, 19], and the three-dimensional Hough transformation (HT) [20]. These techniques enable the precise determination of interplanar spacings and angles between crystallographic features with varying degrees of success. For the work reported here, we have developed an extension of the advanced SDM technique introduced in ref. [19] to enable investigation of the distribution of element-specific inter-atomic separations (e.g. partial pair correlation functions) along particular crystallographic directions. Similar approaches have previously been implemented in the *atom vicinity* algorithm [21], or conventional spatial distribution maps [18], and exploited, for example, to examine site occupancy in intermetallics [22, 23, 24, 25]. However, such a method has never been applied to ordered precipitates in a ternary alloy to determine the specific localization of a specific species.

After casting, the experimental alloy was homogenized for 8 h at 500°C, hot rolled, solution treated for 15 min at 500°C and cold water quenched. The material was subsequently heat treated for 8h at 150°C or 24h at 120°C. To prepare atom probe specimens, the raw material was first cut into 0.4 x 0.4 x 30 mm blanks with a low-speed diamond saw. Needle-shaped specimens were then fabricated by electropolishing in a solution of 25% perchloric acid in glacial acetic acid at 10 – 12 V DC, followed by a second stage of fine polishing under a binocular microscope with 2% perchloric in 2-butoxyethanol at 20 – 25 V DC [26]. Specimens were analyzed in an Imago LEAP 3000X Si under a pulse fraction of 20%. The specimens were maintained at cryogenic temperatures (~40 K) under ultrahigh vacuum conditions of approximately 4.5 $10^{-9}$ Pa. A detection rate of 0.5 – $2x10^{-2}$ ions-per-pulse was maintained through the experiment. Tomographic data containing 12-100 million ions were acquired. Our tomographic reconstructions were generated by using the protocol described in ref. [14], and the reconstruction parameters (i.e. image compression factor $\xi$ and field factor $k_f$) were calibrated using the partial crystallographic information retained within the atom probe data following the protocols introduced in refs. [27, 28]. Data reconstruction and visualization were performed with commercial software (Cameca IVAS).

Typical atom probe tomograms for the respective heat treatments are presented in Figure 1(a–b). In each of these datasets, major crystallographic planes were observable, notably (111) and (022) in the specimen treated for 8h at 150°C, and (002) in the specimen treated for 24h at 120°C. Figure 1 (c–d), provide close-ups of thin slices through the reconstructions presented in Figure 1 (a–b) where atomic planes are imaged illustrating the atomic resolution available. Isoconcentration surfaces, which delineate regions containing more than 8 at.% Li, are included to aid visualization of the δ′ precipitates, which are known to adopt a L1$_2$ (see Figure 1 (e)).

As reported in Table 1, the overall composition measured in the experiment is affected by the *preferential field evaporation* of Li and, to a lesser extent, of Mg. These elements require a lower electric field to be field evaporated than Al [29] and hence may field evaporate under the standing voltage in between the HV-pulses. This prevents the time-of-flight measurement and precludes their identification, which induces an element-specific loss of these atoms. In these studies, preferential field evaporation caused the loss of 15–30% of the Li, with analyses of the material treated for 24h at 120°C being the most affected. The matrix composition for each element and in each dataset was derived from the first nearest-neighbor distribution with the DIAM method described in ref. [30], and values are reported in Table 1. By using the Li 8 at.% isoconcentration surface as a reference proximity histogram were computed [31]. These composition profiles, shown in Figure 2, enable measurement of the composition of the precipitates. The composition of the δ′ precipitates was taken as the average value of the plateau visible in the proximity histograms calculated for more than 100 precipitates. These results are displayed in Figure 2, and the values are reported in Table 1. The size of the precipitates was estimated from the volume embedded by each individual isoconcentration surface, and the assumption that they were approximately spherical, as seems the case from previous transmission electron microscopy experiments [1, 32]. The average size of the precipitates was found to be larger in the material treated for 8h at 150°C than when treated 24h at 120°C, as reported in Table 1. Interestingly, in both cases, there is almost no change in the Mg concentration in the matrix and in the δ′ precipitates. Preferential evaporation of Li and Mg can explain the difference in the measurement of the precipitate composition between the two samples, but also the gap with the expected $Al_3$(Li, Mg) stoichiometry.

To investigate the Mg site-occupancy within the δ′ precipitates, the tomogram for the specimen aged for 8h at 150°C was cropped to retain only the volume encompassing the {022} planes. Likewise, the tomogram from the sample aged for 24h at 120° was cropped to include the region containing the

{002} planes. Subsequently, the 8 at. % Li isoconcentration surfaces were used to generate two subsets of the data: (i) the atoms from the precipitates that were embedded within these surfaces, and (ii) the atoms from the surrounding matrix. Species-specific spatial-distribution maps (SDM) were then computed in both sub-volumes and the resulting histograms plotted in Figure 3. The peak-to-peak distance of Mg-Li distributions, for example, indicate the likely positions of Li atoms relative to a Mg atom along specific crystallographic directions. All peak-to-peak distances in the matrix data of Fig. 3 correspond to a single interplanar spacing of the Al matrix. Significantly, this experimental self-consistency held for the samples analysed in both heat treatment conditions, and in all matrix SDMs and indicates that the species Mg and Li are occupy all sites of the FCC-Al matrix with equal probablity. In comparison, close analyses of the SDMs corresponding to the $\delta'$ precipitate phase revealed a clear change in the periodicity of the Li-Li and Mg-Mg distances. Moreover, the magnitude of the change in periodicity confirms that ordering has taken place, and the structure is consistent with a $L1_2$ structure shown inset in Figure 3. Finally, the peaks associated with Mg appear at the same location as the Li in the $L1_2$ structure, which indicates that most of the Mg atoms partition to the Li sub-lattice within the $\delta'$ precipitates. The exact amount of Li and Mg on the different sites of the L12 structures cannot be precisely determined using the method introduced here due to the filtering of the data based on the isoconcentration surface, which cannot ensure that atoms from the matrix are not also included, which could dramatically affect composition measurements. Other methods for data filtering can be envisaged in the future to try and reveal exactly composition of each sub-lattice.

To investigate the theoretical stability associated with the ordering reaction observed experimentally within $\delta'$ precipitates, we performed all-electron DFT calculations by using the generalized gradient approximation [33] with the $DMol^3$ program package [34]. The wave functions were expanded in terms of a double-numerical quality basis set, with a set of large values of real-space cutoff (Li: 13.76 Bohr; Mg: 12.09 Bohr; and Al: 12.75 Bohr). A reciprocal space of 3x3x3 K-point meshes was employed in the calculations of the 32-atom $Al_3Li$-based supercells. We allowed full atomic relaxation, until the forces on the atoms were less than 0.005 eV/Å. The fully optimized lattice constants for bulk Mg, Al and Li, as well as $Al_3Li$ were in good agreement with experimental values (errors below 1.85%). To evaluate the relative stability of various sub-lattice configurations, the energy required for substitution of Mg on either the Al ($Mg_{Al}$), or Li ($Mg_{Li}$) sub-lattice of the $L1_2$ structure is expressed as $E^f = E_{Mg:Al3Li} - E_{Al3Li} - E_{Mg} + \mu_{Al\ or\ Li}$, where $E_{Mg}$, $E_{Mg:Al3Li}$ and $E_{Al3Li}$ are, respectively, the total energies of bulk Mg, Mg-doped $Al_3Li$, and of the pure $Al_3Li$ reference structure as calculated with the same size supercell. The $\mu_{Al\ or\ Li}$ is the chemical potentials of Al or Li. The chemical potentials $\mu_{Al\ or\ Li}$ depend on the actual environment surrounding the supercell (reservoir). Here, surrounding the precipitate is an Al-rich (above 85 at.%) matrix, hence we used $\mu_{Al} = \mu_{Al\ (bulk)} = E_{Al(bulk)}$, and $\mu_{Li} = \mu_{Al3Li} - 3\mu_{Al\ (bulk)} = E_{Al3Li(bulk)} - 3E_{Al(bulk)}$, and we invoke the relationship $3\mu_{Al} + \mu_{Li} = \mu_{Al3Li}$, assuming both species are in thermal equilibrium with $Al_3Li$ [35]. The obtained formation energies: $Mg_{Al} = 0.43$ eV, and $Mg_{Li} = 0.28$ eV. Similar methods were used to calculate the formation energy of Mg substituting in the pure Al matrix. The value obtained was 0.35 eV, which is intermediate between the energies for Mg substitution within the precipitates. These DFT calculations provide direct insights into the energetic origins of our atomic resolution microscopy observations. Firstly, it is clear that there is a favorable energetic situation when Mg partitions to the Li sites within the δ′ precipitates, as was observed. Secondly, the small difference in the amount of Mg between the matrix and the precipitates can be explained by the relatively small difference in total energy between the insertion of a Mg in the Al-matrix and its most favorable energy state of Mg when substituted for Li in the $L1_2$ structure of the δ′ precipitates.

In summary, we have proposed an original atom probe microscopy approach to identify and characterize ordering within the structure of nanoscale precipitates. Our atomic resolution microscopy experiments, in conjunction with first principles DFT calculations, demonstrate that partitioning of Mg to the Li sub-lattice occurs in L1$_2$ structured δ′ precipitates in an Al–5.18Mg –6.79Li (at. %) alloy aged for 8h at 150°C, and 24h at 120°C. The composition of the Li-rich δ′ precipitates was shown to be very close to the expected Al$_3$(Li, Mg) stoichiometry in both sets of samples, with Li and Mg both more concentrated in the 8h at 150°C. The composition in Mg is almost the same in both the precipitate and matrix phases, and we have demonstrated unequivocally that Mg partitions to the Li sites of the L1$_2$ structure within the δ′ phase precipitates. These observations correspond to the most energetically favorable scenarios, as confirmed by DFT calculations.

## ACKNOWLEDGEMENTS

The authors are grateful for scientific and technical input and support from the Australian Microscopy & Microanalysis Research Facility (AMMRF) at The University of Sydney. Rongkun Zheng is thanked for support and discussions. Joffrey Dremont is thanked for experimental assistance.

## FIGURE AND TABLE CAPTIONS

Table 1: for the two heat treatments: overall composition, composition and size of the δ' precipitates, and matrix composition obtained from the DIAM method [30].

|  | | 8h at 150°C | | | 24h at 120°C | | |
| --- | --- | --- | --- | --- | --- | --- | --- |
|  | element | overall | δ' precipitates | matrix | overall | δ' precipitates | matrix |
| at.% | Li | 5.72 | 17.15 ± 0.54 | 4.67 | 4.67 | 13.97 ± 0.59 | 4.88 |
|  | Mg | 5.53 | 6.43 ± 0.35 | 5.26 | 5.26 | 5.17 ± 0.38 | 5.81 |
|  | Al | 88.75 | 76.15 ± 0.61 | 90.07 | 90.07 | 80.86 ± 0.67 | 89.31 |
|  | radius (nm) | | 3.85 ± 1.37 | | radius (nm) | 2.24 ± 0.86 | |

Figure 1: Typical three-dimensional reconstruction from atom probe analyses of the alloy after (a) 8h at 150°C and (b) 24h at 120°C shown at the same scale. (c) is a close-up showing the (022) atomic planes imaged in the data set displayed in (a) while (d) is a close-up showing the (002) atomic planes imaged in the data set displayed in (b). (e–f) Model view of the fcc-Al matrix and of the $L1_2$ structure. Note that the same colour code is used throughout the article.

Figure 2: Proxigrams computed from the 8 at% Li isoconcentration surface for the material aged (a) 8h at 150°C and (b) 24h at 120°C. Composition profiles of Li and Mg are respectively plotted in pink and green. The dashed lines show the position of the interface at 8 at% Li.

Figure 3: Normalised species-specific spatial distribution maps for the two heat treatments in the matrix (top row) and the precipitates (bottom row). Note the change in the periodicity between the precipitates and the matrix in both cases. The corresponding structure is shown inset in each case.

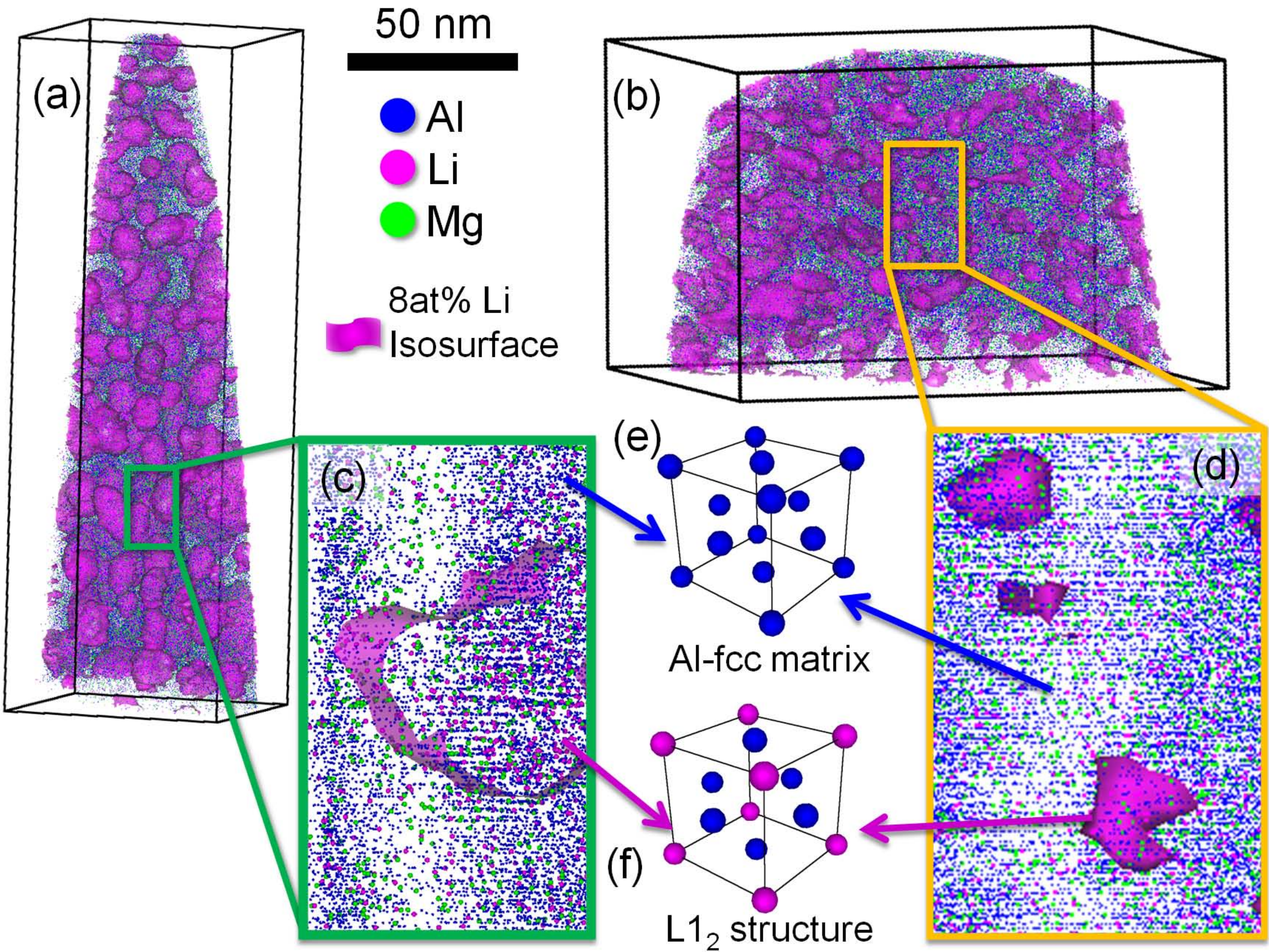

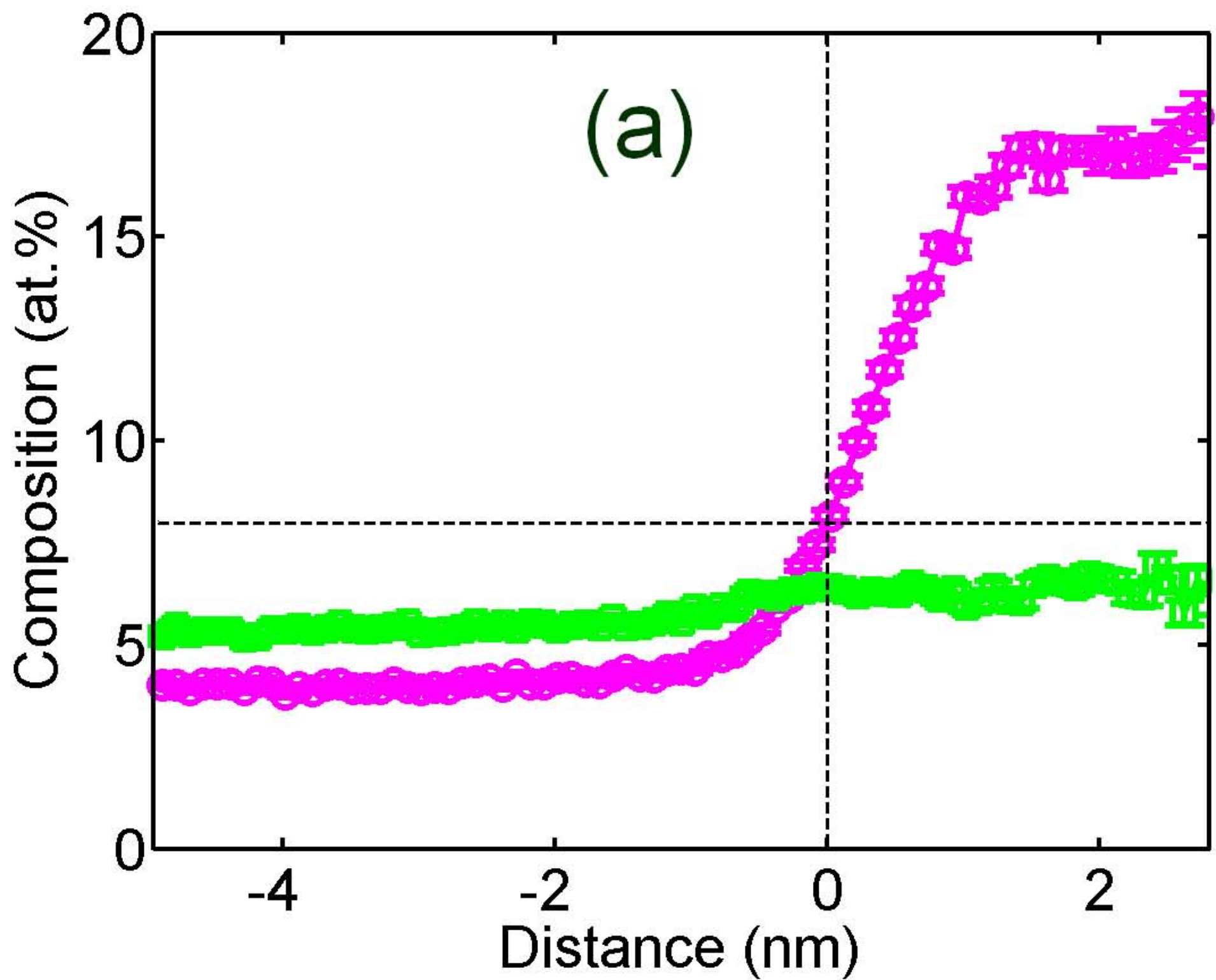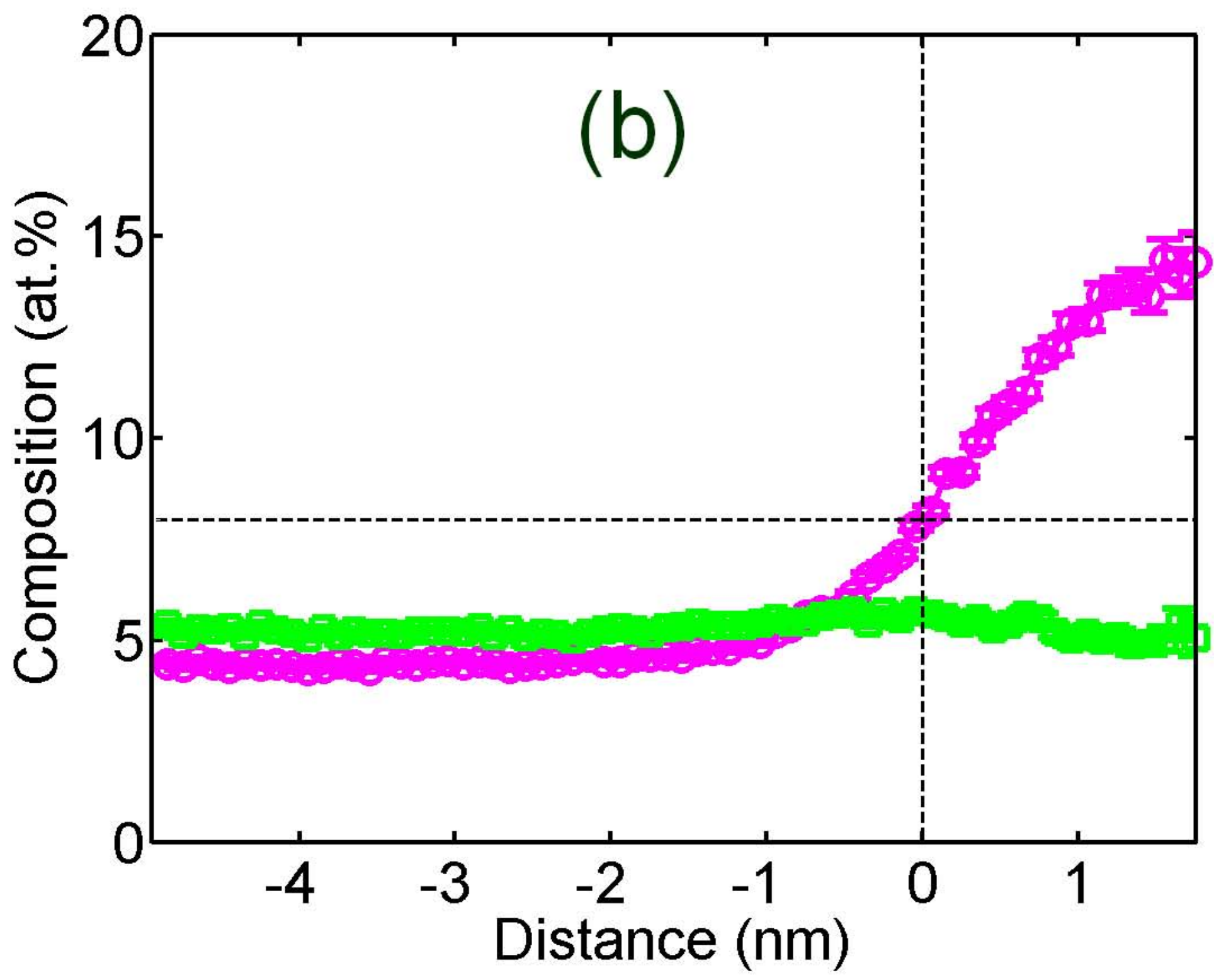

*8h @ 150°C* along the **[011]**  *24h @ 120°C* along the **[001]**

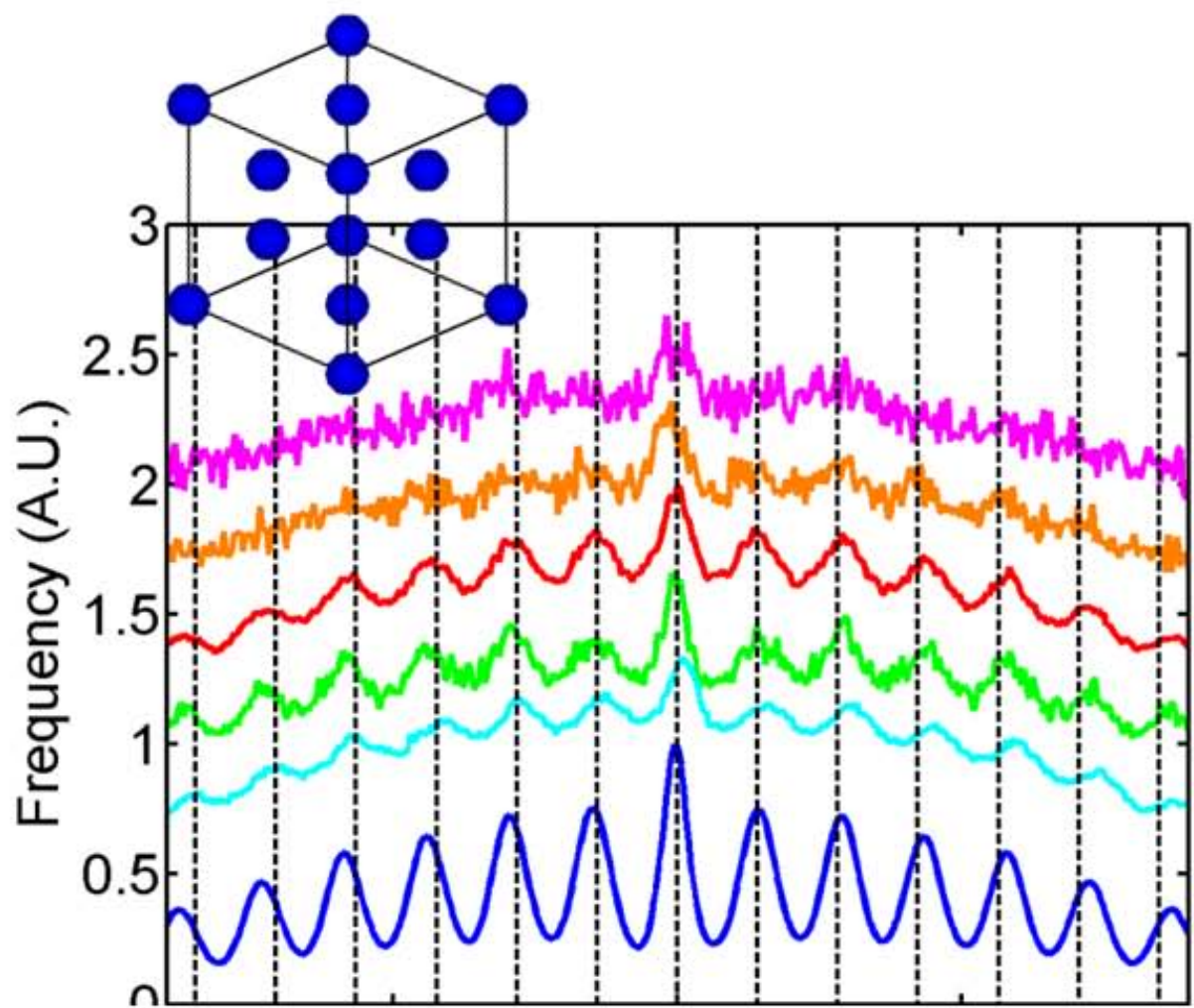
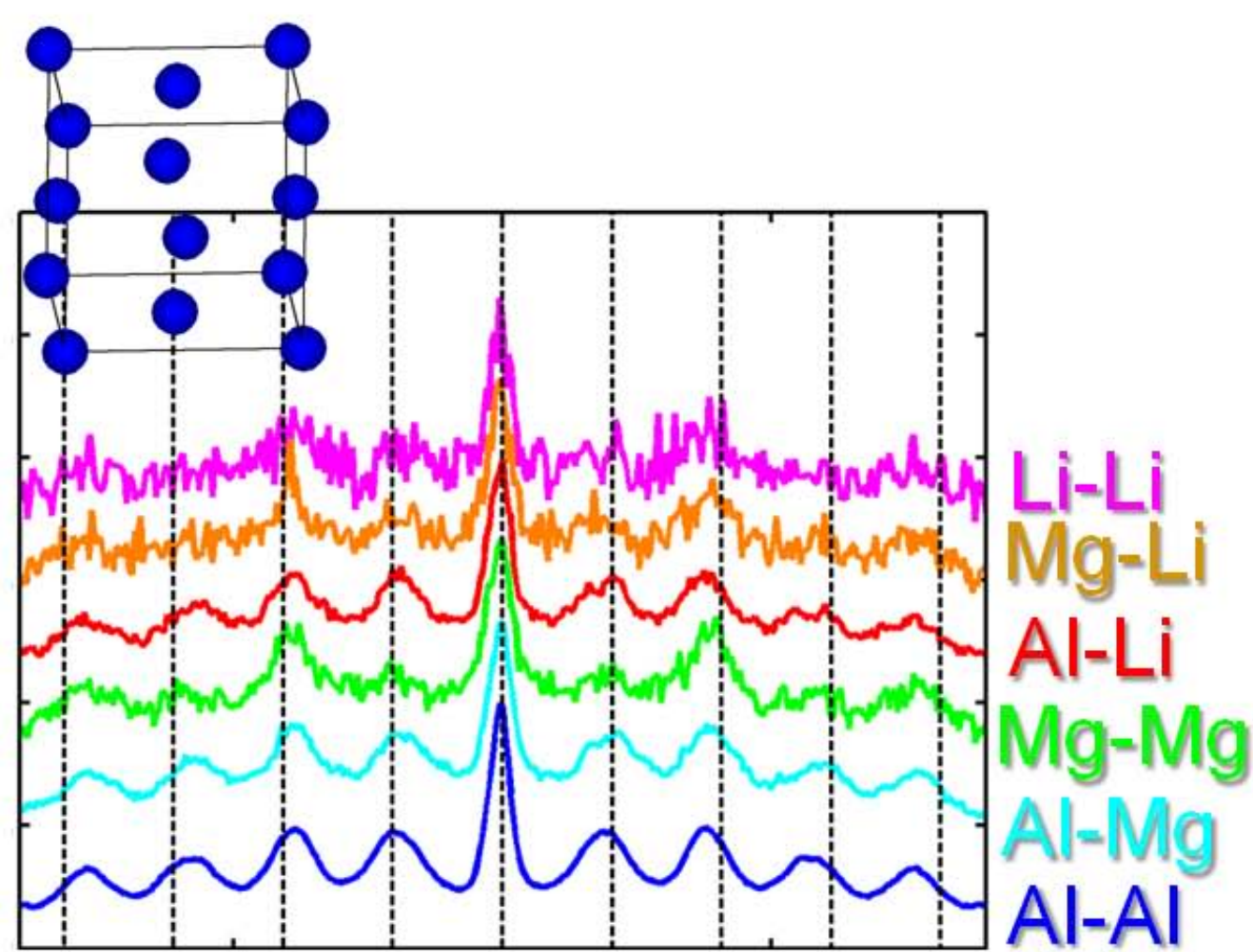
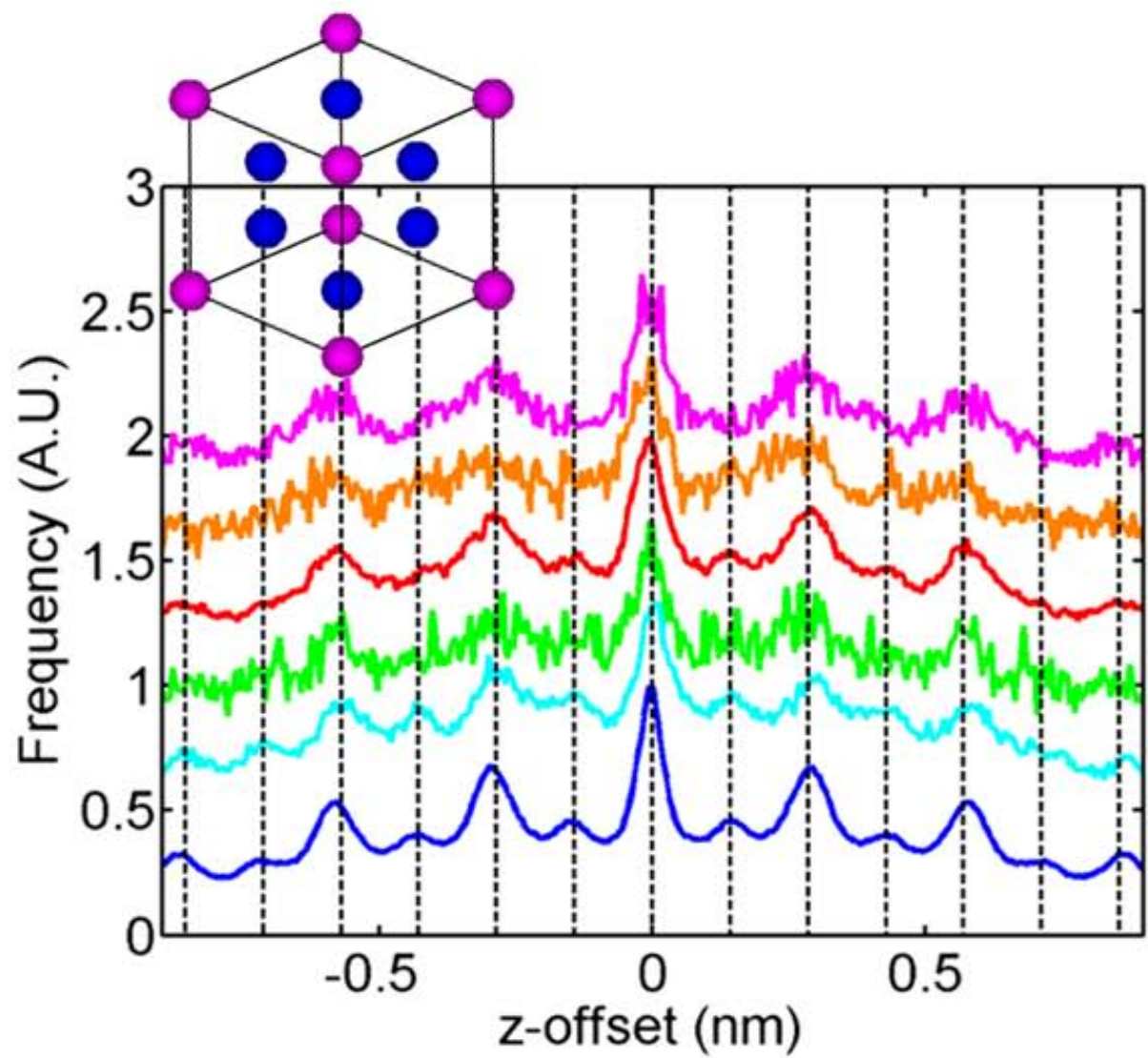
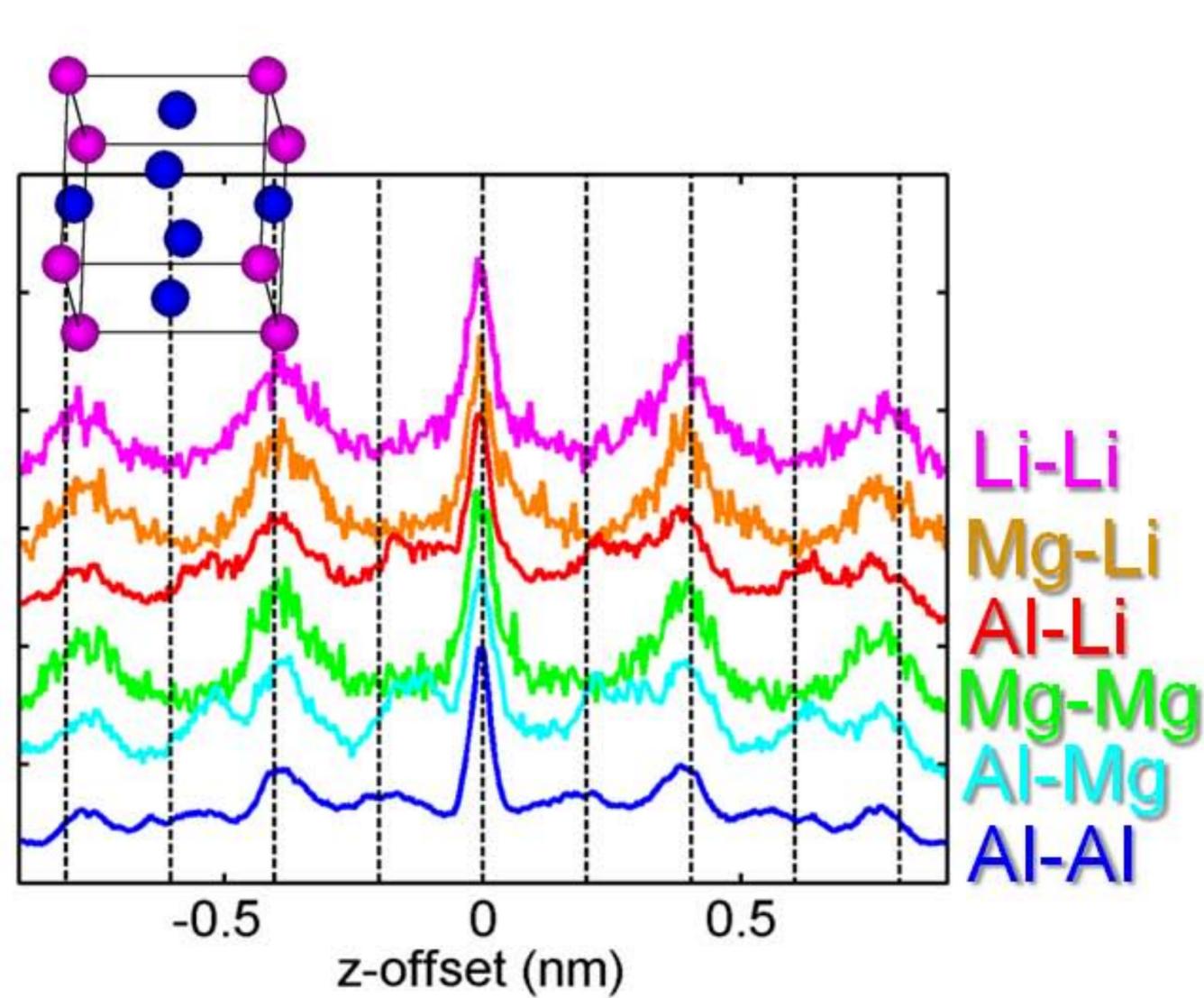